# Multipath Routing With Novel Packet Scheduling Approach In Wireless Sensor Networks

Mary Cherian, T. R. Gopalakrishnan Nair

*Abstract*—Wireless sensor networks sense and monitor real-time events. They supervise a geographic area where a phenomenon is to be monitored. The data in sensor networks have different levels of priority and hence their criticality differs. In order to keep up the real time commitment, the applications need higher transmission rates and reliability in information delivery. In this work we propose a multipath routing algorithm which enables the reliable delivery of data. By controlling the scheduling rate, it is possible to prevent congestion and packet loss in the network. The algorithm provides an efficient way to prevent the packet loss at each node. This results in congestion management in the sensor networks. This protocol prevents packet clustering and provides smoothness to the traffic. Through monitoring and controlling the scheduling rate the flow control and congestion control are managed.

## I. INTRODUCTION

The technology advancements in wireless communications, sensors and embedded systems have made it made possible to implement very large wireless sensor networks. However the inherent characteristics of these networks pose challenges to reliability, information assurance, security and privacy. The ultimate goal of the sensor networks is to obtain meaningful global information from the local information gathered by the individual nodes. The information will become unuseful unless it is reliable, assured and within the dead line [10].

Usually the sensors are deployed in large scale. There are situations where nodes are embedded and they remain unattended. Individual sensors do not have identities and are not aware of the global topology [3]. These sensors happened to have short connecting range and hence they follow multihop communication. In multipath routing, a sensor node should adjust its own data sending rate and the sending rate of its child nodes in a fair and scalable manner [12]. A node can have multiple parents and multiple paths to the sink. But all the nodes may not have multiple paths to the sink. Since the sensor nodes have limited communication range, multiple hopes are required from a source to the sink. Load balancing is very important in wireless sensor networks, since the bandwidth available for each node is limited.

The remaining part of the paper is organized as follows. Section 2 gives an overview on the protocol and section 3 describes the related work. Section 4 illustrates the results and section 5 brings out the conclusions of the work.

## II. MULTIPATH ROUTING TECHNIQUE

Each sensor node initiates multiple flows which can be classified according to their requirements of reliability, throughput, delay and transmission rate. QoS metrics include bandwidth, latency or delay, jitter and throughput [16]. These metrics or their variants will be used for WSNs depending on the application requirement. For a delay-sensitive application, WSNs may also require timely delivery of data. The types of traffic vary from simple query and reply events which are periodic to unpredictable sudden bursts of event messages that are generated by the sensor nodes [2]. The messages are to be delivered within the deadlines otherwise they loose their significance. In this paper, we present an algorithm to minimize the packet loss and hence to obtain the desirable throughput for real-time communication.

Multipath routing involves construction of multiple paths between the source and the destination. We assume that the initial phase of path discovery procedure is already carried out with routing protocols like R2TP [5]. Each sensor node will transmit the data it originates as well as forwards the data passing through the node which is called transit data. Two separate queues are maintained for each type of originating data from the node and three different queues for the transit data. Consider the network model as shown in the fig.1. Network model is aimed, to help a node requesting a certain service to the network, to find the most appropriate route providing the right requested service. Initially a queuing model and the network model are initialized. Every sensor node has equal number and same type of sensors. Source dynamically assigns the priority of the individual application data. For medium access control all nodes are assumed to use MAC protocol like CSMA/CCA [14]. The Fig.1 depicts multipath routing from the node S to the sink.

A prioritizer is provided at the network layer and it classifies the traffic according to the priority of the data and places the data in the appropriate queue. Packets can be prioritized by reading the packet header which includes a priority number for each type of packet. The priority number is assigned at the source end. The data packets are scheduled for transmission based on the priority assigned, by a scheduling unit. It decides the order of service for the data packets. The scheduling unit consists of a software layer which applies the Earliest deadline first (EDF) algorithm

[15].

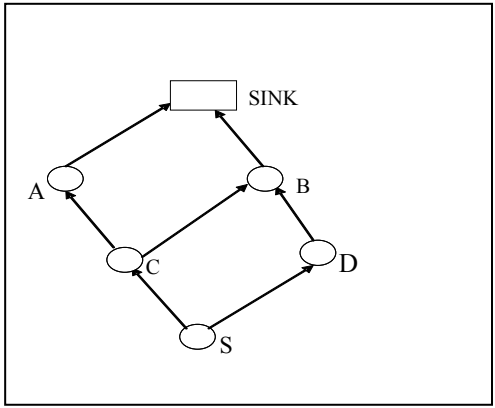

Fig. 1. Network model

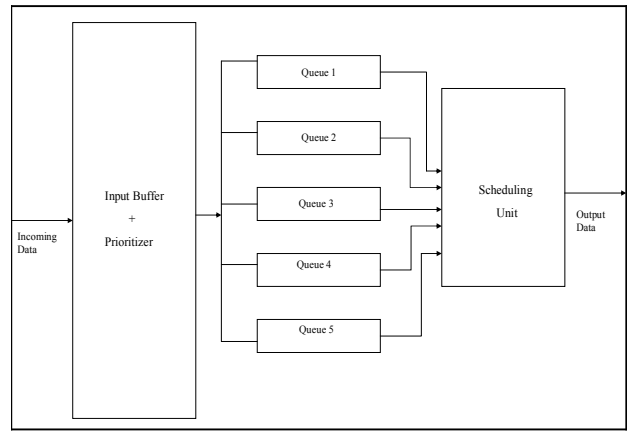

Fig. 2. The Queuing model for a node

EDF algorithm assigns priorities to individual jobs in the tasks according to their absolute deadlines. By controlling the scheduling rate, it is possible to prevent congestion and packet loss in the network.

EDF algorithm will search for the process closest to its deadline. This process is the next to be scheduled for execution. The processor utilization factor,

$$U = \sum_{i=1}^{n} \frac{C_i}{T_i} \leq 1,$$

where the $\{C_i\}$ are the worst-case computation-times of the *n* processes and the $\{T_i\}$ are their respective inter-arrival periods [17].

The Fig.2 depicts the queuing model for a node. There will be inter queue priority and intra queue priority. The queue for transit data has higher priority than the queue for the originating data. Each node checks the priority number and the source address at the header of the packet on the arrival of the packet, orders the packet and puts the packet in the appropriate queue. Once the queue is selected, the packet which has higher priority is selected from the header of the packet by reading the priority number and the EDF algorithm will schedule the transmission. A multipath routing protocol involves the path discovery, traffic distribution and maintenance of the paths [9]. The scheduling rate is denoted by $Sch^i_r$ which gives the number of packets the scheduling unit schedules per unit time from the queues.

For each node with multiple parents, the net scheduling rate of a node is the sum of scheduling rate required for each of the parent node.

Let $O^i_r$ be the originating rate for node i.
$Sch^i_r$ be the scheduling rate for node i and $Sch^{k,i}_r$ be the scheduling rate of parent k of node i.

The scheduling rate of any node will be sum of scheduling rates of all its parent nodes [12].

$$Sch^i_r = \Sigma\ Sch^{k,i}_r \qquad (1)$$

$Sch^i_r$ is the rate at which the scheduling unit schedules packets per unit time from the priority queues.

By controlling the scheduling rate, the flow control and congestion control are done. The scheduling unit sends the packets to the MAC layer which in turn transmits the packets to the network. Initially the scheduling rate of the parent nodes of a node will be distributed. It remains unchanged until the packet service ratio for any parent goes beneath a threshold. When the packet sevice ratio falls below certain value the queue starts building up at the node. Once the queue length crosses the threshold value the packet scheduling rate of the parents should be reduced.

By adjusting the scheduling rate the buffer overflow is avoided. Each node piggybacks the queue length, the packet scheduling rate and the packet service rate in the packet header. Each node also calculates the originating rate of the packets. The originating rate depends on the scheduling rate and the priority of the data[12].

III. RELATED WORK

Earlier circuit-switching networks used alternate path routing to decrease call blocking. Two exchanges use shortest path till the resources are exhausted and afterwards an alternate path is used even though a longer one. Multipath routing for wireless sensor network is an emerging research area. STCP [4] (Sensor Transmission Control Protocol) considered multiple sensing devices in the same node, but does not use multipath routing. Since STCP uses an ACK/NACK based scheme, the delay incurred may not be acceptable for meeting the requirements of deadlines in WSN in critical applications.

In R2TP [5], the packet forwarding is based on the time metric. The algorithm tries to achieve reliability by duplicating the packets in multiple paths. But the paper does not elaborate the queuing model at the nodes or the scheduling strategy of forwarding the packets for meeting the stringent requirements of deadlines in WSN. A Review of Multipath Routing Protocols: From Wireless Ad Hoc to Mesh Networks [9] elaborates the multipath routing techniques for wireless Ad hoc and mesh networks. But the routing protocols for Ad Hoc networks will not suit the sensor networks.

RAP is a real time communication protocol which uses velocity monotonic scheduling (VMS) [2]. Higher priority is assigned for packets which require higher velocity. VMS improves the deadline miss ratio of the WSN. Assumption is that each sensor knows its own location. Velocity is calculated based on the end to end deadlines and the communication distance.

Speed [1] is another real time protocol developed for WSN. Speed and RAP are based on geographic forwarding and are soft real time solutions. But both Speed and RAP does not use multipath routing.

A Survey of transport protocols for wireless sensor network [7] presents a survey of transport protocols for Wireless Sensor Networks (WSNs) and highlights the basic design criteria and challenges of transport protocols which include energy-efficiency, quality of service, reliability, and congestion control. But multipath routing scenarios are not considered.

In Priority based Congestion Control Protocol (PCCP) [8] a node priority based control mechanism has been proposed for WSN. PCCP prioritizes both source and transit traffic but has limitation in handling multiple sensed data within a node.

## IV. SIMULATION RESULTS AND DISCUSSIONS

The simulation determines the threshold value of Packet Service Ratio, Path Length graph showing the average number of hops needed to reach the responded sensor and also Energy Consumption Graph showing Energy consumed by each of the sensors in the network.

The details of simulation parameters are as follows: In an area of 50x50 m$^2$ sensor field, 100 sensors are deployed randomly. Sensors are having a transmission range of 12 m. Number of executions is 2 (service request by each sensor). The maximum Rate adjustment value is 70% and is also assumed that there is no interference from other nodes. The maximum queue length is considered to be 8 packets, with a packet size of 30 bytes.

The lifetime of a wireless sensor network is constrained by the limited energy and processing capabilities of its nodes. To extend the life time of the sensor networks it is very important to have high energy efficiency at all the processing nodes. The Fig.3 below indicates the energy consumption values at the prioritiser, scheduling unit and due to congestion. The energy consumption E is given by

$$|E|^\alpha = O(d^\alpha) \quad [13] \qquad (2)$$

where α is the attenuation factor which can have values from 2 to 5 and d denotes the distance to the receiving node. As we can see from the data, the energy consumption is decreasing

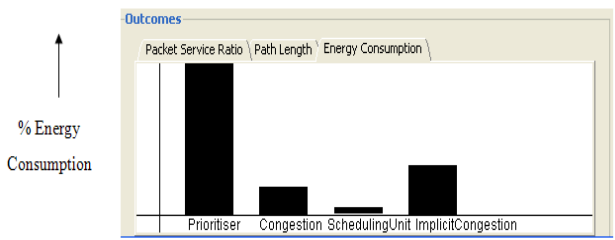

Fig. 3. Energy Consumption

as we apply the algorithm.

TABLE I depicts the energy consumption values of the prioritizer, scheduling unit and due to congestion.

TABLE I: ENERGY CONSUMPTION (JOULES)

| Prioritizer | Congestion | Scheduling Unit | Implicit Congestion |
|---|---|---|---|
| 2068.5 | 452.0 | 560.0 | 1105.0 |
| 2943.0 | 237.5 | 210.0 | 0.5 |
| 1817.0 | 241.5 | 435.0 | 0.5 |
| 723.0 | 0.5 | 87.0 | 112.0 |

**Packet Service Ratio:** Reliability of data in wireless sensor networks depends on the packet reliability which requires the successful reception of the packets at the base station within the specified success ratio. The packet service rate is the inverse of the delay at the sensor node. This delay time will include the time from which the packet has been received at the receive buffer at the node, till the time at which the packet is retransmitted from the node. Packet service ratio, $r_i$ can be used as a measurement to control the scheduling rate at each node i. It is the ratio of average packet service rate denoted by $S_r$ and packet scheduling rate $Sch_r$ in each sensor node i as follows:

$$r_i = S_r^i / Sch_r^i \quad [12] \qquad (3)$$

TABLE II: PACKET SERVICE RATIO

| Packet service Ratio(r) | Packet Drop (%) |
|---|---|
| 0.5 | 80 |
| 1.0 | 78 |
| 1.5 | 70 |
| 2.0 | 64 |
| 2.5 | 62 |
| 3.0 | 60 |

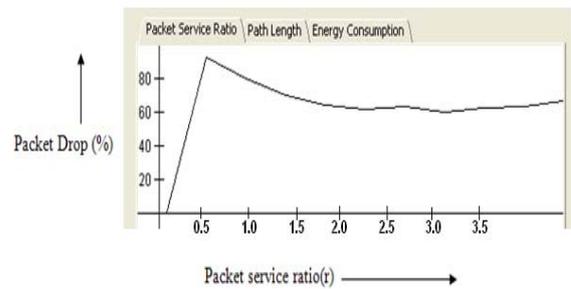

Fig. 4. Packet Service ratio v/s packet drop

As we can see from the table, the packet drop decreases as the packet service ratio increases. A packet service ratio lesser than 0.5, indicates that the scheduling rate requires to be decreased by which the packet drop can be reduced. By reducing the scheduling rate the buffer overflow at the node is controlled [12]. The Fig.4 indicates the packet service ratio to the Packet Drop in the network.

Path Length: Due to the specific restrictions related to wireless sensor networks, the resources consumption saving is a critical issue. Therefore, a shorter path is chosen which involves less number of sensors and consequently, less utilization of resources such as energy or bandwidth.

Path length shows the average length of all the paths found between the sensors in the network. The Table.3 depicts the Path Length for the given number of hops.

Given below is the algorithm for finding the Path Length.

**Path Length Algorithm**
for k=1 to no. of sensors

```
do
S_k is initial sensor
find path from S_k to neighbors
do
for every returned k
do
if (k>current_best_path) then
current_best_path is k
while(timeout does not expire)
```

Success Rate : As we increase the packet service ratio, we can observe that the success rate of packet delivery at the destination increases. The graph in Fig.5 illustrates this. The loss of packets can be due to buffer overflow, network congestion or link level collisions.

TABLE III: PATH LENGTH

| No. of Hops | Path length |
|---|---|
| 1 | 1.2 |
| 2 | 2.0 |
| 3 | 2.43 |
| 4 | 2.43 |

Bur*st data management* : Sudden changes in the total volume of traffic at any node can deteriorate the performance of transmission[23]. In multipath routing the incoming large flows of data will be transferred by the sensor nodes in the outgoing multiple paths. Thus it reduces the effects of bursty traffic [22]. Since this is a multipath routing algorithm, load balancing is achieved by data traversing the available multiple paths from the source to the destination.

## V. OPTIMALITY OF THE ALGORITHM

The Principle of Optimality states that " An optimal policy has the property that whatever the initial state and initial decisions are, the remaining decisions must constitute an optimal policy with regard to the state resulting from the first decision" [18],[19]. At a particular time the value of a decision problem is decided with respect to some initial choices and the remaining values of the problem results from the initial choices.

In our algorithm we start with a small initial value for the packet service ratio, and we observe that the packet drop is increasing. Here the scheduling rate and the data output rate of the sensor nodes are high, which results in buffer overflow and packet loss. By increasing the packet service ratio the packet drop decreases. This is achieved by reducing the scheduling rate. Also if the packet service ratio of any of the parent nodes falls below, subsequently the sensor node will reduce its scheduling rate. Finally the desired scheduling rate is achieved according to the dynamic conditions of the network.

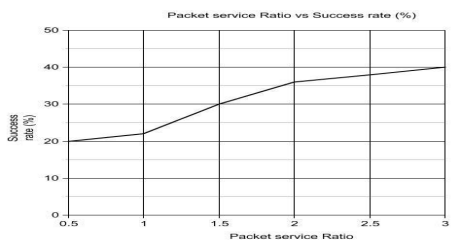

Fig. 5.Packet Service ratio v/s success rate

The scheduling unit uses the EDF algorithm for scheduling the packets. With a set of n independent tasks with random arrival times as in the case of sensor nodes, the EDF algorithm is optimal in minimizing the maximum lateness. HORN'S algorithm [21] states that "Given a set of n independent tasks with arbitrary arrival times, any algorithm that at any instant executes the task with the earliest absolute deadline among all the ready tasks is optimal with respect to minimizing the maximum lateness".

Let $m_{EDF}$ be the schedule obtained by the EDF algorithm for a schedule m of algorithm B. The schedule m has time slices of one unit of time each.

Let m(t) identifies the task executing in the slice [t,t+1] e(t) identifies a task that is ready at time t, which has the earliest deadline. $t_e(t)$ is the time at which the next slice of task e(t) begins its execution in the current schedule. Interchanging the position of m(t) and e(t) cannot increase the maximum lateness[20].

In the algorithm used by Dertouzos [20] to transform any schedule m into an EDF schedule, for each time slice t, the algorithm verifies whether the task m(t) scheduled in the slice t is the one with the earliest deadline, e(t). If not a transposition takes place and the slices at t and $t_e$ are exchanged. The slice of task e(t) is expected at time t, while the slice of task m(t) is postponed at time $t_e$. After each transposition the maximum lateness cannot increase. So the algorithm is optimal.

## VI. CONCLUSION

This paper has discussed the implementation of multipath routing protocol in a sensor network environment. The algorithm prevents the packet loss at each node by adjusting the queue length. This results in congestion management in the sensor networks. This protocol prevents packet clustering and provides smoothness to the traffic [11]. Through monitoring and controlling the scheduling rate, the flow control and congestion control are managed. Over-feedback to multiple parental nodes could call for a situation where, a certain amount of required data may tend to loose very close to the hop points of origination. Since our focus is on burst data scenario salvation, we expect incremental data buffering facility of the nodes would accommodate the burst data on the propagating paths, without resulting in a major data loss. The algorithm is a dynamic one since the output rate of the sensor nodes are adjusted depending on the network conditions. We will be further working on an admission control algorithm which will ensure congestion control in case of bursty traffic. This work may also be extended to predictive congestion control for wireless sensor networks. Scheduling rate control can be performed dynamically based on a prediction model.

ACKNOWLEDGMENT

Authors thank Revansiddappa,, Dr. Ambedkar Institute of Technology for the support for the required simulation in the laboratory.

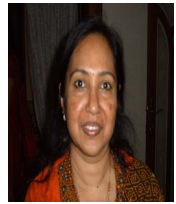

**Mary Cherian** has 27 years of experience in professional field spread over Education, Research and Industry. She holds B.E degree in electronics and communication from Kerala University, India (1983) and M.Tech.in computer science and engineering from visvesvaraya technological university Belgaum, India (2005). She started her career in 1984 as research engineer in O/E/N India and has worked in the field of engineering and software in industries like Kerala state electronics development corporation, Keltron controls, Electronics Research and Development Centre, ABB and Chemtrols Software private limited in capacities of System Engineer , System Manager and Director. Later, she concentrated on education and contributed in academic field for the last 10 years in India and abroad educating pupil in the field of science and technology especially in Computer Science and Engineering. Currently she is working as Assistant Professor in the department of Computer Science in Dr. Ambedkar Institute of Technology , Bangalore , India. She has publications in national and international conferences. Her areas of interests include Computer networks, Sensor Networks, Real time routing protocols, and Cognitive routing . Prof Cherian has membership of IACSIT and life membership of professional bodies CSI, ISTE, IE and IETE.

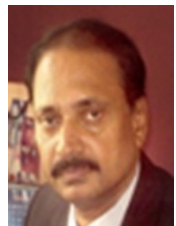

**T. R. Gopalakrishnan Nair** has 30 years of experience in professional field spread over Education, Research and Industry. He holds degrees M.Tech. from I.I.Sc., Bangalore, India and Ph.D. in computer Science, from Kerala University, India. He is currently the Vice President (Research and Development) of one the largest education centers, DS Institutions in Bangalore, India. He started his career in Electronics Research and Development Centre, Trivandrum, India where he was instrumental in developing various pioneering research products in the field of computers and software. Later, as the head of Advanced Simulation Activities in Indian Space Research Organization, his areas of research were Critical Real-Time Systems, Inertial Navigation and Guidance systems, Launch Vehicle Technology, High speed computing and Launch Vehicle simulations. Later, he concentrated on education and contributed extensively in academic field for the last 14 years in India and abroad educating pupil in the field of science and technology especially in Computer Science and Engineering. He is also a consultant to many industries. He authored and published about 110 papers in these multidisciplinary fields and he promotes cross domain fusion of knowledge. He has several patents in the field of computers and communications. He has authored several book chapters at international levels and delivered keynote and invited lectures. Several volumes of his papers and keynotes are brought out, in addition to proceedings that he edited. His areas of interests include Computer networks, Cognitive routing, Software Engineering, Bio-Informatics, AI & Robotics and Signal and Image Processing.

**Dr. Nair** is a senior member of IEEE for last two decades and a member of various other societies like ACM (USA). He is usually invited by various universities at national and international level as well as professional societies for delivering lectures on frontier areas. He is a member of the director boards of few companies operating in India. His articles on current thoughts and lectures on education, technology and business are published by leading Indian publications. He has acted as the chair of many international conferences including the 'Euro-India Research conference –2008' held in Bangalore. He was the chairman of the Expert group of Innovation & Education ( 40 International members) for United Nations Global Alliance for ICT and Development. He is the Chief Editor of Journals "InterJRI Science and Technology and InterJRI Computer Science and Networking" published by Interline Publishers. In 1992, he received the National Technology award PARAM from the advisor to Prime Minister, for developing the parallel computing flight simulation systems. He received the TeamTech Foundation Award for 'Excellence in Education and Research' in 2009.